\documentclass[pra,twocolumn, aps, 10pt, superscriptaddress]{revtex4-2}
\usepackage[english]{babel}
\usepackage{textcomp,xcolor,natbib}
\usepackage{graphicx}
\usepackage{amsmath, bbm}
\usepackage{latexsym}
\usepackage{amsfonts}   
\usepackage{amssymb}
\usepackage{array}      
\usepackage{epsfig}
\usepackage{txfonts}
\usepackage{xcolor,braket}
\usepackage[normalem]{ulem}
\usepackage{soul}
\usepackage{dsfont}
\usepackage{xfrac}  
\usepackage{relsize}
\usepackage{enumitem}
\usepackage{dcolumn}
\usepackage{bm}
\usepackage{ulem}
\usepackage{color}
\definecolor{lavender}{rgb}{0.71, 0.49, 0.86}
\definecolor{verdepino}{rgb}{0.0, 0.6, 0.2}

\usepackage{hyperref}

\begin{document}

\title{Floquet time-convolutionless master equation for non-Markovian driven quantum systems}
\author{Pietro Marco Follia}
\affiliation{Dipartimento di Fisica ``Aldo Pontremoli", Università degli Studi di Milano, via Celoria 16, 20133 Milan, Italy}
\affiliation{Istituto Nazionale di Fisica Nucleare, Sezione di Milano, via Celoria 16, 20133 Milan, Italy}

\author{Heinz-Peter Breuer}
\affiliation{Institute of Physics, University of Freiburg, 
Hermann-Herder-Stra{\ss}e 3, D-79104 Freiburg, Germany}
\affiliation{EUCOR Centre for Quantum Science and Quantum Computing,
University of Freiburg, Hermann-Herder-Stra{\ss}e 3, D-79104 Freiburg, Germany}
\author{Bassano Vacchini}
\affiliation{Dipartimento di Fisica ``Aldo Pontremoli", Università degli Studi di Milano, via Celoria 16, 20133 Milan, Italy}
\affiliation{Istituto Nazionale di Fisica Nucleare, Sezione di Milano, via Celoria 16, 20133 Milan, Italy}

\date{\today}

\begin{abstract}
We study the dynamics of open quantum systems driven by an external time-periodic force. Combining Floquet theory and the time-convolutionless projection operator technique we derive a time-local quantum master equation which exactly takes into account the periodic driving, while treating the system-environment interaction within second order in the coupling strength without performing the Markov approximation. The resulting equation of motion for the reduced density 
matrix is called Floquet time-convolutionless master equation. Employing the example of the driven spin-boson system, we demonstrate that this master equation is capable of describing strong
non-Markovian effects, while yielding the Floquet-Lindblad master equation in the Markovian limit. A characteristic feature of memory effects in such driven dissipative systems is the emergence of sharp peaks of the trace-distance based non-Markovianity measure as a function of the driving amplitude, which can be traced back to quasienergy crossings leading to almost 
decoherence-protected subspaces through a quasienergy-induced dissipative decoupling mechanism. 
\end{abstract}

\maketitle

\section{Introduction}

The realistic description of any quantum platform requires accounting for its unavoidable interaction with the surrounding environment. Within the framework of open quantum systems, this coupling typically triggers dissipation and decoherence, leading to the loss of quantum features~\cite{Breuer2002,Vacchini2024}. However, in  strong-coupling regimes or in the presence of structured spectral densities, a bidirectional flow of information between system and environment can emerge. This phenomenon, known as quantum memory effects or non-Markovianity~\cite{Rivas2014a,Breuer2016a,Chruscinski2022a}, plays a pivotal role both in the foundations of physics and in the development of quantum technologies, where manipulating memory effects is crucial to extending coherence times. In recent years, the use of external time-periodic fields has emerged as a powerful tool to engineer and control the dynamics of open quantum systems. Most standard approaches to the dissipative dynamics of periodically driven systems rely on a Markovian description based mainly on the Floquet-Lindblad master equation~\cite{Blumel1991a,Breuer2002, Grifoni_1998}. However, these frameworks fail to correctly reproduce the dynamics near quasienergy degeneracies, where the standard secular approximation breaks down. Although refined approaches based on the partial secular approximation have been developed to address this issue~\cite{Farina_2019a,Mozgunov2020,one_over_f_noise}, they are inherently designed to maintain a time-independent or Markovian structure, thus failing to capture relevant non-Markovian effects. To overcome these limitations, other extensions of the description for periodically driven open quantum systems have been proposed~\cite{Hausinger2010a,Mori2023a, Mickiewicz2026}.
The non-trivial role of quasienergy degeneracies in determining non-Markovianity has been recently established in Ref.~\cite{Follia2026a}, where we investigated a periodically driven spin-boson model at finite temperature using the numerically exact Hierarchical Equations of Motion (HEOM). This work revealed an unexpected, sharp peak-like structure in the trace-distance-based non-Markovianity measure when considered as a function of the driving amplitude, showing that quasienergy crossings lead to a strong enhancement of the system's relaxation times and hence the  emergence of long-lived memory effects. However, the microscopic interpretation we  provided in~\cite{Follia2026a}  was based on the Markovian Floquet-Lindblad master equation, thus only providing indirect evidence. In this manuscript, we derive a non-Markovian master equation for dissipative two-level systems under general periodic driving which provides the most convenient formalism to understand the relationship between quasienergy degeneracies and non-Markovianity. Our approach starts from the time-convolutionless projection operator technique~\cite{Shibata1977a,Chaturvedi1979a,Breuer2002}, and systematically exploits Floquet theory to express the system's coupling operator in the Floquet basis~\cite{Breuer2002,Blumel1991a}. We then show that, under certain assumptions, the mathematical and physical structure of this equation is profoundly modified by the presence of quasienergy degeneracies, giving rise to a mechanism that  we call \emph{quasienergy-induced
dissipative decoupling}. Finally, we demonstrate the crucial importance of this modification by studying its impact on the non-Markovianity of a linearly driven spin-boson system, proving that the proposed Floquet time-convolutionless framework successfully captures the genuine non-Markovian features of the dynamics.

The paper is organized as follows: In Sec.~\ref{sec:ftcl} we introduce the Floquet time-convolutionless master equation that arises from combining the standard time-convolutionless projection operator technique with Floquet theory, highlighting the role played by the Floquet-Fourier components in determining the reduced dynamics, as well as the crucial structural change in the expression of the master equation that takes place in the presence of quasienergy degeneracies. Sec.~\ref{sec:tls} is focused on adapting the general theoretical framework to the specific case of a two-level system interacting with a bosonic environment. In Sec.~\ref{sec.Theo_NM} we review the definition of non-Markovianity that we employ, together with the corresponding measure, while Sec.~\ref{sec:nm} investigates the memory effects of the considered dynamics. Finally, we comment on our results and point to future developments in Sec.~\ref{sec:ceo}.






\section{Floquet time-convolutionless master equation}
\label{sec:ftcl}
In this Section, we  briefly introduce the time-convolutionless projection operator technique, a perturbative strategy that allows us to derive the equations of motion for the reduced dynamics of a quantum system interacting with a general bath. We will apply this perturbative technique to the situation in which the system of interest is exposed to a periodic driving, taking full advantage of Floquet theory. The derivation is carried out in the interaction picture with respect to the full Hamiltonian of the driven system. We build upon the work of Bl\"umel \textit{et al.}~\cite{Blumel1991a} and exploit the Floquet decomposition of the system propagator to express the system coupling operator in the Floquet basis, providing a natural and compact representation of the dissipative dynamics which immediately highlights the crucial effects of quasienergy degeneracies.
We focus in particular on the second-order time-convolutionless master equation for a driven two-level quantum system, which we will call Floquet time-convolutionless master equation. Importantly, the exploitation of the Floquet basis and the associated quasienergies in introducing the interaction picture allows us to put into evidence a peculiar structural change of the expression of the master equation in the presence of degeneracy of quasienergies. Indeed, the interplay between driven system Hamiltonian and coupling to the environment is fully encoded in the Floquet-Fourier coefficients that determine the relevance of the different contributions in the master equation. In the case of degeneracy an appropriate change of basis leads to a significant simplification of both the coherent and dissipative contributions to the master equation.
The general treatment remains valid for an arbitrary periodic driving. We finally provide the Bloch-vector representation of the obtained reduced dynamics, which, besides its independent interest, will be especially convenient for the treatment of non-Markovian effects in Sect.~\ref{sec:nm}.


\subsection{Interaction picture with Floquet states}
Let us consider the following Hamiltonian describing a periodically driven open quantum system
\begin{equation}
    H_{\mathrm{SB}}(t) = H_{S}(t)\otimes \mathbb{I} + \mathbb{I}\otimes H_{B} + H_I,
\end{equation}
where $H_B$ is the free bath Hamiltonian, the interaction term is $H_\mathrm{I} = \sum_{\alpha}X_{\alpha}\otimes B_{\alpha}$ and the reduced system Hamiltonian is periodically dependent on time:
\begin{equation}
    H_S(t+T) = H_{S}(t), \quad T = 2 \pi /\omega.
\end{equation}
The starting point of the time-convolutionless projection operator technique is the von Neumann equation in the interaction picture~\cite{Breuer2002},
\begin{equation}
\label{eq:von_neumann}
    \frac{d}{dt}\rho_\mathrm{SB}(t) = -i[H_I(t), \rho_\mathrm{SB}(t)],
\end{equation}
where
\begin{equation}
    H_I(t) =\sum_{\alpha}X_{\alpha}(t)\otimes B_{\alpha}(t) =\sum_{\alpha}U_\mathrm{S}^\dagger(t)X_{\alpha}U_\mathrm{S}(t)\otimes U^\dagger_\mathrm{B}(t)B_{\alpha}U_\mathrm{B}(t),
\end{equation}
with $U_{S,B}(t) = \mathcal{T}\exp\left(-i\int_0^tdt'H_{S,B}(t')\right)$.
The periodic time dependence of the system Hamiltonian allows us to employ the Floquet decomposition of the reduced system unitary time-evolution operator
\begin{equation}
    \label{eq:floquet_theorem}
        U_S(t) = \sum^{N}_{n=1} \ket{u_n(t)} \bra{u_n(0)} \mathrm{e}^{-i\varepsilon_n t},
\end{equation}
where $\ket{u_n(t)}$ are the Floquet states, which form a complete orthonormal basis of the Hilbert space at each fixed time $t$ and are periodic with the same period $T$ as the Hamiltonian. The real quantities $\varepsilon_n$ denote the corresponding quasienergies~\cite{Breuer1988a,Breuer1988b,Zeldovich1967a,Shirley1965a,Ritus1967a,Sambe1973a,Goldman2014a,Holthaus2015a}, while $N$ is the dimension of the reduced system Hilbert space. Within this framework, the interaction-picture coupling operators can be expressed as
\begin{align}
\label{eq:G_coupling_op_floquet}
    X_{\alpha}(t) &=  U_\mathrm{S}^\dagger(t)X_{\alpha}U_\mathrm{S}(t) \nonumber \\
    &=  \sum_{n,n'}\langle u_{n'}(t)|X_\alpha|u_n(t)\rangle\, \mathrm{e}^{-i(\varepsilon_n-\varepsilon_{n'})t} \, |u_{n'}\rangle\langle u_{n}|.
\end{align}

For the sake of simplicity, let us discuss in more detail the case of a two-level system in the following. In this case, the quasienergies can be chosen to satisfy $\varepsilon_1+\varepsilon_2=0$, so that, upon defining $\varepsilon\equiv\varepsilon_1$, one has $\varepsilon_2=-\varepsilon$~\cite{Holthaus2015a}. 
The Floquet states at the reference time $t=0$, denoted by $|u_i\rangle\equiv|u_i(0)\rangle$, provide a basis in the Hilbert space that allows us to introduce the following operators
\begin{align}
\label{eq:floquet_basis_zpiumeno}
    \Sigma_z &=  \frac{1}{\sqrt{2}}\left(|u_1\rangle\langle u_1| - |u_2\rangle\langle u_2|\right), \nonumber \\
    \Sigma_+ &=  |u_1\rangle\langle u_2| , \nonumber \\
    \Sigma_- &=  |u_2\rangle\langle u_1| ,
\end{align}
which together with the identity yield a basis of operators.
In this basis, Eq.~\eqref{eq:G_coupling_op_floquet} for a two-level system becomes
\begin{align}
\label{eq:TLS_coupling_op_floquet}
    X_{\alpha}(t) = c_{+,\alpha}(t) \mathrm{e}^{2i\varepsilon t}\, \Sigma_+ +  c_{-,\alpha}(t) \mathrm{e}^{-2i\varepsilon t}\,\Sigma_- +  c_{z,\alpha}(t)\,\Sigma_z,
\end{align}
where we have defined
\begin{equation}
\label{eq:c_coefs_generic}
    c_{z,\alpha}(t) = \sqrt{2}\langle u_1(t) |X_{\alpha}|u_1(t)\rangle, \quad c_{+,\alpha}(t) = \langle u_1(t) |X_\alpha|u_2(t)\rangle ,
\end{equation}
and $c_{-,\alpha}(t)=c^*_{+,\alpha}(t)$.
The system coupling operators $X_\alpha$ have been assumed, without loss of generality, to be Hermitian and traceless so that $\langle u_1(t) |X_\alpha|u_1(t)\rangle = - \langle u_2(t) |X_\alpha|u_2(t)\rangle$ and $\langle u_1(t) |X_\alpha|u_2(t)\rangle^* =\langle u_2(t) |X_\alpha|u_1(t)\rangle$.

For simplicity, and without loss of generality, henceforth we will consider the case of a single coupling operator on the system side, i.e., $H_I = X \otimes \sum_{\alpha}B_\alpha$, and consequently drop the subscript $\alpha$ when referring to the system's coupling operator.
\subsection{Floquet-Fourier coefficients}
Exploiting the periodicity of the Floquet states it is convenient to express the amplitudes $c_{\pm}(t)$ and $c_{z}(t)$ as Fourier series
\begin{equation}
\label{eq:general_fourier_floquet_coeff}
    c_{\pm} (t) = \sum_{n=-\infty}^{+\infty} c^{n}_\pm \mathrm{e}^{i\omega nt}, \quad c_z(t) =  \sum_{n=-\infty}^{+\infty} c^{n}_z \mathrm{e}^{i\omega nt},
\end{equation}
where the Fourier coefficients satisfy the relations $\big(c_+^n\big)^* = c^{-n}_-$ and $\big(c^n_z\big)^* = c_z^{-n}$. 
We refer to $c_{\pm}^{n}$ and $c_z^{n}$ as \emph{Floquet-Fourier coefficients}.
In general, the computation of the Floquet-Fourier coefficients requires solving the dynamics of the isolated system, which can be achieved analytically only for a restricted class of models, while numerical approaches are available in general~\cite{Follia2026a,Hausinger2010a}. 
\subsection{Master equation for two-level system}
We now proceed to derive the time-convolutionless master equation using the decomposition in the Floquet basis of the interaction picture coupling operator  given in Eq.~\eqref{eq:TLS_coupling_op_floquet}.
Following standard treatments, we employ the time-convolutionless projection operator technique~\cite{Breuer2002}, which provides time-local master equations while retaining  memory effects. Starting from the von Neumann equation~\eqref{eq:von_neumann} and assuming a factorized initial state $\rho_{SE}(0)=\rho_S(0)\otimes\rho_E$,
the reduced dynamics takes the form
\begin{equation}
    \frac{d}{dt}{\rho}_S(t)=\mathcal{K}(t){\rho}_S(t),
\end{equation}
where $\mathcal{K}(t)$ is the time-convolutionless generator that admits a perturbative expansion in powers of the system-bath coupling strength
\begin{equation}
    \mathcal{K}(t)=\sum_{n=1}^{\infty}\mathcal{K}_n(t).
\end{equation}
Assuming $\mathrm{Tr}_E \{ B_\alpha\rho_E \}=0$
the first-order contribution vanishes identically and truncating the expansion at second order 
the obtained master equation determined by $\mathcal{K}_2(t)$ takes the explicit expression
\begin{equation}
\label{eq:TCL2}
    \frac{d}{dt}{\rho}_S(t) = -\int_0^t ds\,
    \mathrm{Tr}_E
    \left[H_I(t),\left[H_I(s),\rho_S(t)\otimes\rho_E\right]\right].
\end{equation}
Finally, substituting the Floquet basis decomposition of the system coupling operator given in Eq.~\eqref{eq:TLS_coupling_op_floquet}
we obtain
\begin{align}
\label{eq:Floquet_TCL_zpm}
  \frac{d}{d t} \rho_S (t) =  &- i \left[H_\mathrm{LS}(t), \rho_S (t) \right]
  \nonumber\\
  &  + \sum_{i, j = z, +, -} a_{ij} (t) \left[ \Sigma_i \rho_S (t)
  \Sigma_j^{\dagger} - \frac{1}{2} \{ \Sigma_j^{\dagger} \Sigma_{i}, \rho_S (t) \}
  \right] ,
\end{align}
where the Lamb shift Hamiltonian is given by
\begin{equation}
\label{eq:abstrac_LS_Hamilton}
     H_\mathrm{LS}(t) = h_z (t) \Sigma_z +\frac{1}{2 i} \left( h_+ (t) {\Sigma_+}  - h.c. \right),
\end{equation}
and $a_{ij}(t)$ are the elements of a generalized Kossakowski matrix whose explicit expression is given below. 
Equation~\eqref{eq:Floquet_TCL_zpm} will be referred to as  \emph{Floquet time-convolutionless} (Floquet-TCL) master equation throughout this work.
For the sake of compactness, we introduce the notation 
\begin{equation}
\label{eq:gamma}
  \Gamma \left[ {f }  (t), g  (t) \right] =  \int_0^t ds \left[ f (t)
  g^{\ast} (s)C^{\ast} (t - s) + f  (s) g^{\ast} (t)C(t -
  s) \right] ,
\end{equation}
and
\begin{equation}
\label{eq:pi}
  \Pi \left[ {f }  (t), g  (t) \right] = \int_0^t d s \left[ f
  (t) g^{\ast} (s) - f  (s) g^{\ast} (t)\right] \mathrm{Re}\{C (t - s)\},
\end{equation}
where $C(t-s)=\sum_{\alpha,\alpha'}\langle B_\alpha (t) B_{\alpha'} (s) \rangle $ is the environmental correlation function.
Using this notation the Kossakowski coefficients appearing in the Floquet-TCL master equation~\eqref{eq:Floquet_TCL_zpm} can be expressed as
\begin{align}
\label{eq:Kossa_zpm}
  a_{zz} (t) & =  \Gamma [c_z (t), c_z (t)], \nonumber\\
  a_{z \flat} (t) & =  \Gamma [c_z (t), c_\flat (t)\mathrm{e}^{\flat 2i\varepsilon t}],  \nonumber\\
  a_{\flat \sharp} (t) & =  \Gamma [c_\flat (t) \mathrm{e}^{\flat 2i \varepsilon t}, c_\sharp (t) \mathrm{e}^{\sharp 2i\varepsilon
  t}],
\end{align}
where $\flat,\sharp\in\{+,-\}$, so that  $a_{- +} (t)=a_{+ -}^* (t)$,
while the coefficients appearing in the expression of the Lamb shift Hamiltonian \eqref{eq:abstrac_LS_Hamilton} read 

\begin{align}
\label{eq:LS_Hamiltonian}
  h_z (t) & =  \frac{i}{\sqrt{2}} \Pi [c_+ (t) \mathrm{e}^{+2i\varepsilon t}, c_+ (t)
  \mathrm{e}^{+2i\varepsilon t}], \nonumber\\
  h_+ (t) & =  
  \sqrt{2} \Pi [c_z (t), c_- (t) \mathrm{e}^{-2i\varepsilon t}].
\end{align}
We emphasize that the Floquet-TCL master equation Eq.~\eqref{eq:Floquet_TCL_zpm} applies to a general two dimensional system, considering the second-order approximation in the time-convolutionless projection operator technique. Higher-order approximations, as well as distinct perturbative approaches such as the Nakajima-Zwanzig projection operator technique~\cite{Nakajima1958a,Zwanzig1960a} can be considered in the same spirit, exploiting the Floquet states to introduce the interaction picture and expressing the effect of coupling to the bath in terms of quasienergies and Floquet-Fourier coefficients.

\subsection{Change of basis and degenerate case}
With the aim of highlighting how the Floquet-TCL master equation is modified in the presence of a quasienergy degeneracy, i.e., when $\varepsilon=0$, it is convenient to express it in the alternative operator basis
\begin{align}
\label{eq:floquet_basis_zyx}
    \Xi_z &=  \frac{1}{\sqrt{2}}\left(|u_1\rangle\langle u_1| - |u_2\rangle\langle u_2|\right), \nonumber \\
    \Xi_y &=  \frac{i}{\sqrt{2}}\left(-|u_1\rangle\langle u_2| +|u_2\rangle\langle u_1|\right), \nonumber \\
    \Xi_x &=  \frac{1}{\sqrt{2}}\left(|u_1\rangle\langle u_2| + |u_2\rangle\langle u_1|\right),
\end{align}
in terms of which Eq.~\eqref{eq:Floquet_TCL_zpm} can be written as
\begin{align}
\label{eq:Floquet_TCL_zyx}
  \frac{d}{d t} \rho_S (t) =  &- i \left[H_\mathrm{LS}(t), \rho_S(t) \right]
  \nonumber\\
  &  + \sum_{i, j = z, y, x} b_{ij} (t) \left[ \Xi_i \rho_S (t)
  \Xi_j^{\dagger} - \frac{1}{2} \{ \Xi_j^{\dagger} \Xi_{i}, \rho_S (t) \}
  \right], 
\end{align}
where the coefficients $b_{ij}(t)$ are the elements of the Kossakowski matrix in this basis, which will be denoted by $\mathcal{B}(t)$.
Its matrix elements can be expressed in terms of the $a_{ij}(t)$ coefficients of Eq.~\eqref{eq:Kossa_zpm} as
\begin{align}
\label{eq:kossa_zyx}
    b_{zz}(t) &=  a_{zz}(t), \nonumber \\
    b_{yy}(t) &=  \tfrac{1}{2}\left(a_{++}(t)+a_{--}(t)\right)- \mathrm{Re}\{a_{+-}(t)\}, \nonumber \\
    b_{xx}(t) &=  \tfrac{1}{2}\left(a_{++}(t)+a_{--}(t)\right) +\mathrm{Re}\{a_{+-}(t)\}, \nonumber \\
   b_{zy}(t) &=  -\tfrac{i}{\sqrt{2}}\left(a_{z+}(t)-a_{z-}(t)\right), \nonumber \\ 
    b_{zx}(t) &=  \tfrac{1}{\sqrt{2}}\left(a_{z+}(t)+a_{z-}(t)\right), \nonumber \\
    b_{yx}(t) &=  \tfrac{i}{2}\left(a_{++}(t)-a_{--}(t)\right)-\mathrm{Im}\{a_{+-}(t)\},
\end{align}
and the others follow from the fact that $\mathcal{B}(t)$ is Hermitian. Finally, in this basis, the Lamb shift Hamiltonian can be re-written as
\begin{equation}
   \label{eq:LS_hamiltonian_zyx}
    H_{\mathrm{LS}}(t) = h_z (t) \Xi_z  +  \tfrac{1}{\sqrt{2}}\mathrm{Im}\{h_+(t)\}\Xi_x+ \tfrac{1}{\sqrt{2}}\mathrm{Re}\{h_+ (t)\} \Xi_y . 
\end{equation}

We now consider the case in which the quasienergies of the isolated system are degenerate. Substituting $\varepsilon = 0$ in~\eqref{eq:Kossa_zpm} and assuming  $c_{+}(t)\in\mathbb{R}$, we immediately obtain thanks to Eq.~\eqref{eq:gamma}
\begin{equation}
   \label{eq:deg1}
    a_{z+}^{\mathrm{D}}(t) = a_{z-}^{\mathrm{D}}(t), 
\end{equation}
as well as
\begin{equation}
   \label{eq:deg2}
    a_{++}^{\mathrm{D}}(t) = a_{--}^{\mathrm{D}}(t) = a_{+-}^{\mathrm{D}}(t),
\end{equation}
so that, according to Eqs.~\eqref{eq:kossa_zyx}, the Kossakowski matrix simplifies to
\begin{equation}
\label{eq:kossa_deg}
\mathcal{B}^\mathrm{D}(t) = \begin{pmatrix}
    a_{zz}(t) & 0 & \sqrt{2}a^{\mathrm{D}}_{z+}(t)\\
    0  & 0 &0 \\
    \sqrt{2}\left(a^\mathrm{D}_{z+}(t)\right)^{*} & 0 &2a^\mathrm{D}_{++}(t)
\end{pmatrix},
\end{equation}
where the superscript D denotes the degenerate case and the ordering of the components is taken to be $z$, $y$, $x$. Likewise, according to~\eqref{eq:pi}, \eqref{eq:LS_Hamiltonian} and~\eqref{eq:LS_hamiltonian_zyx} the Lamb-shift Hamiltonian simplifies to
\begin{equation}
        H^{\mathrm{D}}_{\mathrm{LS}}(t) =\tfrac{1}{\sqrt{2}}\mathrm{Re}\{h_+^{\mathrm{D}} (t)\} \Xi_y.
\end{equation}
The coherent contribution to the dynamics is confined to the $y$ direction, whereas the dissipative contribution is restricted to the $x$--$z$ plane. We call this mechanism quasienergy-induced dissipative decoupling. 
Recalling its very definition that follows from  Eq.~\eqref{eq:c_coefs_generic}
\begin{equation}
\label{eq:c_piu_again}
    c_{+}(t) = \langle u_1(t) |X|u_2(t)\rangle ,
\end{equation}
we see that the function $c_{+}(t)$ can always be chosen to be real if the matrix associated to the coupling operator $X$ in the Floquet basis determined by $H_S(t)$ is real. More generally, this structure of the dissipator at the quasienergy degeneracy also emerges when the imaginary part of $c_{+}(t)$ is small, i.e., $\mathrm{Re}\{c_{+}(t)\}\gg\operatorname{Im}\{c_{+}(t)\}$. Indeed, in this regime the coefficients $b_{yy}(t)$, $b_{yz}(t)$, and $b_{yx}(t)$ become small and can be neglected. Under the same hypothesis, the coherent contribution is also simplified, since the terms $h_z(t)$ and $h_x(t)$ can likewise be neglected. A symmetric situation arises when $c_{+}(t)$ is almost purely imaginary. In this case, the coherent contribution is confined to the $x$ direction, while the dissipative part is confined to the $y$--$z$ plane.
\subsection{Bloch equation}
It is convenient to express the reduced state in the operator basis~\eqref{eq:floquet_basis_zyx} through its Bloch representation,
\begin{equation}
    \rho_S(t) = \frac{1}{2}(\mathbb{I}_2+ \boldsymbol{x}(t)\cdot \boldsymbol{\Xi}),
\end{equation}
where the Bloch vector components are given by $x_i(t) = \mathrm{Tr}_{S}\{\rho_S(t)\Xi_i\}$. In this representation, the master equation~\eqref{eq:Floquet_TCL_zyx} takes the form
\begin{equation}
    \dot{\boldsymbol{x}}(t) = M(t)\boldsymbol{x}(t) + \boldsymbol{c}(t),
\end{equation}
where $M(t)$ is a $3 \times 3$ real matrix called \emph{dynamical matrix} and $\boldsymbol{c}(t)$ is a three dimensional vector. It is always possible to decompose the dynamical matrix in its symmetric and antisymmetric parts,
\begin{equation}
    M(t) = M^{\mathrm{S}}(t) + M^\mathrm{A}(t)
\end{equation}
with
\begin{equation}
    M^\mathrm{S}_{ij}(t) = \mathrm{Re}{\{b_{ij}(t)\}} -\mathrm{Tr}\{\mathcal{B}(t)\}\delta_{ij}
\end{equation}
and
\begin{equation}
    M^{\mathrm{A}}_{ij}(t) = - \sum_{k}\varepsilon_{ijk} h_k(t).
\end{equation}
Finally, the vector $\boldsymbol{c}(t)$ can be interpreted as a time-dependent translation with components
\begin{equation}
    c_k(t) = 2\sum_{i,j}\varepsilon_{kij}\mathrm{Im}\{\mathcal{B}(t)_{ij}\}.
\end{equation}
The degenerate case exhibits a remarkable simplification of the Bloch-vector equation as a direct consequence of the modified structure of the dissipator (see Eq.~\eqref{eq:kossa_deg}). In particular, the dynamical matrix reduces to
\begin{equation}
\label{eq:dynamical_deg}
M^{\mathrm{D}}(t) = 
\setlength{\arraycolsep}{1.5pt}
    \begin{pmatrix}
        -b^{\mathrm{D}}_{xx}(t) & 0 &\mathrm{Re}\{b^{\mathrm{D}}_{zx}(t)\} -h_y(t)\\
        0 &-b^{\mathrm{D}}_{zz}(t)-b^{\mathrm{D}}_{xx}(t) & 0 \\
        \mathrm{Re}\{b^\mathrm{D}_{zx}(t)\}+h_y(t)&0&-b^{\mathrm{D}}_{zz}(t)
    \end{pmatrix}
\end{equation}
where as in~\eqref{eq:kossa_zyx} we order the component as $(z,y,x)$.
It follows immediately from the direct sum structure of this matrix that the $y$ component of the Bloch vector evolves independently of the $x$--$z$ components, which remain coupled to each other through both the coherent and dissipative contributions. Therefore, if the initial state of the system lies in the $x$--$z$ plane, the dynamics remains confined to this two-dimensional subspace of the Bloch sphere.
As we show below, by evaluating the non-Markovianity of the dynamics, this significant modification of the Floquet-TCL master equation structure can have important consequences on the dynamics of the open system.

\section{Driven damped two-level system}
\label{sec:tls}
We now demonstrate the validity of the derived master equation by focusing on periodically driven spin-boson systems. After specifying the bath through its spectral density, we consider three representative cases: the undriven two-level system, the two-level system under rotating-wave approximation (RWA) driving, also known as circular driving, and the linearly driven system. For the first case, we show that the Floquet-TCL master equation indeed reduces to the standard second-order time-convolutionless master equation.
\subsection{Spectral density and correlation function}
We consider an open system coupled to a bath of harmonic oscillators with Hamiltonian $H_B = \sum_n\omega_n a^{\dagger}_n a_n$ by the linear interaction term
    \begin{equation}
    \label{eq:spin_boson_interaction_term}
       H_{I} = \sigma_x\otimes\sum_n g_n(a^\dagger_n+a_n),
    \end{equation}
    where  $a_n^\dagger$ and $a_n$ denote the bosonic creation and annihilation operators associated with the harmonic oscillator of frequency $\omega_n$. The coefficients $g_n$ describe the distribution of the couplings between the system and the different harmonic modes.
The bath correlation function is given by
    \begin{equation}
        C(t) = \sum_n|g_n|^2\langle (a_n(t)+a^\dagger_n(t))(a_n(0  )+a^\dagger_n(0))\rangle_{\rho_B},
    \end{equation}
which, by assuming that the bath is initially in the thermal state $\rho_B = \mathrm{e}^{-\beta H_B}/\mathcal{Z}$, can be expressed in terms of the spectral density of the bath according to
    \begin{equation}
    \label{eq:Fluct_diss_theo}
  C (t)  =  \frac{1}{2\pi}\int_0^{+ \infty} d \varpi J (\varpi) \left[ \coth \left(
  \frac{\beta \varpi}{2} \right) \cos (\varpi t) - i \sin (\varpi t) \right].
    \end{equation}
In this work we will assume a spectral density of the Lorentz-Drude form,
    \begin{equation}
    \label{LD_Spectral}
        J(\varpi) = \alpha\varpi\frac{\Lambda}{\varpi^2+\Lambda^2},
    \end{equation}
    where $\Lambda$ represents the cut-off frequency and $\alpha$ the coupling strength between system and bath.
    This widely used expression is particularly suitable for the present work as it allows for the correlation function to be computed analytically. The calculations leading to the explicit expression of the correlation function are rather lengthy and are therefore detailed in Appendix~\ref{Appendix:Correlation Function}.

\subsection{Undriven spin-boson system}
We first consider an undriven system with time-independent Hamiltonian. In this case, the Floquet decomposition of the time-evolution operator~\eqref{eq:floquet_theorem} reduces to
\begin{align}
\label{eq:unitary_undriven}
    U^{\mathrm{UN}}_S (t) = \sum_i|e_i\rangle\langle e_i| \mathrm{e}^{-iE_it},
\end{align}
where ${|e_i\rangle}$ denotes the eigenbasis of the system Hamiltonian and $E_i$ are the corresponding eigenenergies, while the superscript UN refers to the undriven case. In this case, the Floquet states are replaced by the eigenstates of the system Hamiltonian, and the quasienergies by the corresponding eigenenergies.
By taking the Hamiltonian
\begin{equation}
    H^{\mathrm{UN}}_\mathrm{S} = \frac{\omega_0}{2}\sigma_z
\end{equation}
the decomposition~\eqref{eq:unitary_undriven} becomes
\begin{equation}
    U^{\mathrm{UN}}_\mathrm{S} (t) = \mathrm{e}^{-i\omega_0t/2}|e\rangle\langle e| +\mathrm{e}^{+i\omega_0t/2}|g\rangle\langle g|,
\end{equation}
where $|e\rangle, |g\rangle$ are the eigenstates of $\sigma_z$.
The system operator appearing in the interaction term~\eqref{eq:spin_boson_interaction_term} can be written in interaction picture according to Eq.~\eqref{eq:G_coupling_op_floquet} as
\begin{equation}
    \sigma^{\mathrm{UN}}_x(t) = \left(U^{\mathrm{UN}}_S(t)\right)^\dagger \sigma_xU^{\mathrm{UN}}_S(t) = \mathrm{e}^{+i\omega_0t}\Sigma_+ + \mathrm{e}^{-i\omega_0t}\Sigma_-
\end{equation}
with $\Sigma_+ = |e\rangle\langle g|$ and $\Sigma_- = |g\rangle\langle e|$ coinciding with the standard ladder operators. In this case the coefficients are 
\begin{equation}
    c^{\mathrm{UN}}_{\pm} (t) = \mathrm{e}^{\pm i \omega_0t}, \quad c^{\mathrm{UN}}_{z} (t) =0. 
\end{equation}
By substituting this expression and the explicit form of the correlation function~\eqref{eq:explicit_corr_function} in Eqs.~\eqref{eq:Kossa_zpm} and~\eqref{eq:LS_Hamiltonian} it straightforward to compute the master equation for the undriven spin-boson system, that is is well known and has been used in~\cite{NM_sb_any} to study non-Markovianity of the model.
\subsection{Spin-boson system with circular driving}
We now consider the situation in which the system Hamiltonian is given by
\begin{equation}
    H^{\mathrm{RWA}}_S(t) = \frac{\omega_0}{2}\sigma_z + \frac{\Omega}{2}\left(\mathrm{e}^{-i\omega t }\sigma_+ + \mathrm{e}^{i\omega t}\sigma_-\right),
\end{equation}
which describes a two-level system under driving within the RWA.
In this case the quasienergies can be computed analytically as 
\begin{equation}
    \varepsilon = \frac{1}{2}\sqrt{\left(\omega_0-\omega\right)^2+\Omega^2}
\end{equation}
together with the coefficients in Eq.~\eqref{eq:c_coefs_generic} 
\begin{align}
\label{eq:coeff_RWA}
    c^{\mathrm{RWA}}_z (t)&= \sqrt{2} \frac{\Omega}{\sqrt{(\omega - \omega_0)^2 +\Omega^2}} \cos (\omega t), \nonumber \\
    c^{\mathrm{RWA}}_+ (t) &= \frac{\omega - \omega_0}{\sqrt{(\omega - \omega_0)^2
  + \Omega^2}} \cos (\omega t) + i \sin (\omega t) .
\end{align}
By substituting Eq.~\eqref{eq:coeff_RWA} and the analytic expression of the bath correlation function~\eqref{eq:explicit_corr_function} into Eqs.~\eqref{eq:gamma} and \eqref{eq:pi}, one obtains the Floquet-TCL master equation for the RWA-driven spin-boson model. This system has been studied in the standard time-convolutionless formalism in~\cite{Haikka_2010}.
\subsection{Linearly driven spin-boson system}
Finally, we consider the linearly driven spin-boson model with Hamiltonian
\begin{equation}
    H^{L}_\mathrm{S} (t) = \frac{\omega_0}{2}\sigma_z + \Omega\cos\left(\omega t\right)\sigma_x.
\end{equation}
In this case, although the isolated dynamics can be solved analytically in terms of special functions, the resulting expressions generally rely on series expansions~\cite{Xie_2010a,Schmidt_2021a, Ma_2007a,Follia2026b}. 
Furthermore, analytical approaches based on Van Vleck perturbation theory are available~\cite{Hausinger2010a}, but they are not well suited to the present analysis.
We therefore compute both the quasienergies and the Floquet-Fourier coefficients numerically by standard techniques.
For simplicity, we assume a driving frequency close to resonance, $\omega\sim\omega_0$, and restrict our attention to the case in which $\Omega \gtrsim  \omega$, where the Floquet-Fourier coefficients can be accurately approximated by~\cite{Follia2026a}
\begin{equation}
    c^n_{+} = \delta_{n,0}, \quad c^n_{z} = \tfrac{1}{5}(\delta_{n,1}+\delta_{n,-1}).
\end{equation}
Substituting these coefficients into Eq.~\eqref{eq:general_fourier_floquet_coeff} yields
\begin{equation}
\label{eq:coeff_linearly_driven}
    c_{\pm} (t) = 1, \quad c_z(t) = \tfrac{2\sqrt{2}}{5}\cos(\omega t),
\end{equation}
so that in particular $c_{+} (t)$ is well approximated by a real quantity.
Finally, using Eq.~\eqref{eq:coeff_linearly_driven} together with the expression for the bath correlation function~\eqref{eq:explicit_corr_function} to compute the Kossakowski coefficients~\eqref{eq:Kossa_zpm} and the Lamb shift Hamiltonian~\eqref{eq:LS_Hamiltonian} yields the Floquet-TCL master equation for the linearly driven spin-boson model. The explicit expressions for the Kossakowski coefficients and the coherent contribution are rather lengthy and are therefore reported in Appendix~\ref{Appendix:explicit_coefficients}.
%

In the following sections, we will first briefly introduce the definition of Markovianity based on distinguishability and the corresponding measure for non-Markovianity based on the trace distance~\cite{Breuer2009b}. Finally, we will study the non-Markovianity of this linearly driven spin-boson system using the Floquet-TCL master equation, with particular emphasis on the role of quasienergy degeneracies.

\section{Dynamical matrix and distinguishability revivals}
\label{sec.Theo_NM}
In this work the definition proposed in~\cite{Breuer2009b} for Markovianity, which is based on the notion of information backflow, is employed.
The central quantity of this approach is the distinguishability between two quantum states of the reduced system $\rho^A$ and  $\rho^B$ quantified by means of the trace distance~\cite{QInfo,QCmp_QInfo}:
    \begin{equation}
        D(\rho^A, \rho^B) = \frac{1}{2}\mathrm{Tr} \, \big| \rho^A-\rho^B \big|,
    \end{equation}
    where $|A| = \sqrt{A^\dagger A}$. 
    If the dynamics of the open system is described by the family 
    $\{ \Phi_t \mid t\geq 0 \}$ of quantum dynamical maps $\Phi_t$, then 
    for any pair of initial states $\{\rho^A,\rho^B\}$ the time evolution of the trace distance is given by
    \begin{equation}
        D_t\big(\rho^A, \rho^B\big)  = D\big(\Phi_t[\rho^A], \Phi_t[\rho^B]\big).
    \end{equation}
    A monotonically decreasing trace distance, i.e., $ \dot {D}_t \leq 0$, implies that the two states are becoming increasingly less distinguishable, which is interpreted as a unidirectional flow of information from the system to the bath. This clearly defines a Markovian dynamics as the information initially contained in the reduced system is lost in the environment and never retrieved. Conversely, an increase of the trace distance over time, i.e. $\dot{D}_t > 0$, is interpreted as information flowing from the environment into the system. This phenomenon, known as information backflow, is regarded as the characteristic feature of non-Markovian dynamics~\cite{Breuer2009b,Megier2021a}.
Based on this interpretation, a quantitative measure of non-Markovianity for the 
process described by the maps $\Phi_t$
is defined by
    \begin{equation}
    \label{eq:NM}
        \mathcal{N}[\Phi] = \max_{\rho^A,\,\rho^B} \int_{\dot{D}_t > 0} \!\!\! dt \, \dot{D}_t ,
    \end{equation}
where the maximum is taken over all possible orthogonal pairs of initial states.
For a two-level system, as considered in this manuscript, the trace distance between two states reduces to the Euclidean distance between their Bloch vectors,
\begin{equation}
    D(\rho_S^A(t), \rho_S^B(t)) = \frac 12 \|\boldsymbol{x}_A(t)-\boldsymbol{x}_B(t)\|.
\end{equation}
In this setting, it is convenient to introduce the difference vector 
\begin{equation}
    \boldsymbol{d}(t) = \frac{1}{2}\left[\boldsymbol{x}_A(t)-\boldsymbol{x}_B(t)\right].
\end{equation} 
It is well known that, for a two-level system, the pair of initial states maximizing the non-Markovianity measure \eqref{eq:NM} can be chosen to be pure and orthogonal, i.e., antipodal on the Bloch sphere~\cite{Wissmann2012a, non_markov_2015}. Denoting the initial Bloch vectors as $\boldsymbol{x}_0 = \boldsymbol{x}_\mathrm{A}(0) =-\boldsymbol{x}_\mathrm{B}(0)$, the difference vectors evolves according to the homogeneous linear equation
\begin{equation}
\label{eq:dist_dyn}
    \frac{d}{dt}\boldsymbol{d}(t) = M(t)\boldsymbol{d}(t),
\end{equation}
with initial condition $\boldsymbol{d}(0) =\boldsymbol{x}_0$.
The definition of Markovianity is equivalent to the requirement that the norm $\|\boldsymbol{d}(t)\|$ is monotonically non-increasing.
Accordingly, the non-Markovianity measure~\eqref{eq:NM} can be expressed as
\begin{equation}
    \label{eq:NM_bloch}
        \mathcal{N}[\Phi] = \max_{\boldsymbol{d}_0\in S^2} \int_{\frac{d}{dt}\|\boldsymbol{d}(t)\| > 0} \!\!\!\!\!\! dt \, \frac{d}{dt}\|\boldsymbol{d}(t)\| ,
\end{equation}
where $S^2$ denotes the Bloch sphere.
This formalism provides an efficient framework to compute the non-Markovianity of the dynamics associated with the Floquet-TCL master equation, both analytically and numerically, and will be used in the following section.

\section{Non-Markovian peaks in the linearly driven spin-boson system}
\label{sec:nm}
We first recover the Markovian Floquet-Lindblad master equation as the long-time, period-averaged limit of the Floquet-TCL master equation. This procedure allows us to find the damping rates associated to the dynamics in the Markovian limit.
We then address the study of non-Markovianity of the linearly driven spin-boson model, 
showing the impact of quasienergy degeneracy on  non-Markovianity. 

\subsection{Connection to Floquet-Lindblad master equation}
We identify the limit in which the Floquet-TCL master equation derived in this manuscript reduces to a  master equation in Floquet-Lindblad form~\cite{Breuer2002}, which was used in Ref.~\cite{Follia2026a} to analyse the emergence of non-Markovianity peaks in correspondence with the quasienergy degeneracy, that arise for specific values of the driving strength. For simplicity, we restrict the derivation to the degenerate case of the linearly driven spin-boson model, although the same procedure can be straightforwardly extended to the non-degenerate case.
The first step consists in taking the long-time limit of the Kossakowski coefficients $a_{ij} (t)$ of Eq.~\eqref{eq:Floquet_TCL_zpm}, whose exact expression for the considered model is given in Appendix~\ref{Appendix:explicit_coefficients}.
We recall that in the degenerate case, according to Eq.~\eqref{eq:kossa_deg}, the only relevant coefficients are those given in
Eqs.~\eqref{eq:coeff_zz}, \eqref{eq:azp_dege} and \eqref{eq:appapm_dege}.
This long-time limit corresponds to a Markov approximation, in which the relaxation time of the reduced system is assumed to be much longer than the bath correlation time. In practice, this amounts to neglecting all exponentially decaying contributions by replacing them with their asymptotic values. 
Performing this limit the relevant Kossakowski coefficients become
\begin{align}
\overline{a_{zz}^{\mathrm{D}} (t)}^{\infty} = &  \frac{4}{25} \cos (\omega t) \Bigg\{
- \frac{4 \alpha}{\beta \Lambda} \sum_{k = 1}
  \frac{\Lambda^2}{\Lambda^2 - \nu_k^2} \frac{\nu_k
  \omega}{\omega^2 + \nu_k^2} \sin (\omega t)
  \nonumber \\
  & 
  + J (\omega) \left[ \coth \left( \frac{\beta \omega}{2} \right) \cos
  (\omega t) 
  +\cot \left( \frac{\beta \Lambda}{2} \right) \sin (\omega t)\right]  \Bigg\}, \nonumber \\
\overline{a_{z \pm}^{\mathrm{D}} (t)}^{\infty} = &  \frac{1}{5\sqrt{2}} \left\{\alpha\cos
  (\omega t) \left( 3\cot \left( \frac{\beta \Lambda}{2} \right) -\frac{4}{\beta\Lambda} +i\right)
  \right. \nonumber\\
  &  - J (\omega) \left( i - \cot \left( \frac{\beta \Lambda}{2} \right)
  \right) \sin (\omega t) \nonumber\\
  &  - J (\omega) \left( i \frac{\Lambda}{\omega} - \coth \left(
  \frac{\beta \omega}{2} \right) \right) \cos (\omega t) \nonumber\\
  &  - \left. \frac{4 \alpha}{\beta \Lambda} \sum_{k = 1}
  \frac{\Lambda^2}{\Lambda^2 - \nu_k^2} \frac{\nu_k \omega}{\nu_k^2 +
  \omega^2} \sin (\omega t) 
  \right\},
  \nonumber\\
  \overline{a_{\pm \pm}^{\mathrm{D}} (t)}^{\infty} = & \lim_{\omega\to 0} J(\omega)(1+2n(\omega)) = \frac{\alpha}{\beta \Lambda},
\end{align}
where we have introduced the Bose-Einstein distribution $n(x)=1/(\mathrm{e}^{\beta x}-1)$ and $\mathrm{D}$ denotes, as before, the degenerate case.
We finally average the Kossakowski coefficients over one period of the drive, $T= 2\pi/\omega$. This procedure can be interpreted as a coarse-grained description of the dynamics and is commonly associated with the secular (or rotating-wave) approximation. Using the identities $\langle\cos(\omega t)\sin(\omega t)\rangle_T = 0$ and $ \langle\cos^2(\omega t)\rangle_T =\langle\sin^2(\omega t)\rangle_T =1/2$, we obtain for the Kossakowski coefficients of the Floquet-Lindblad master equation the simple expressions:
\begin{align}
\overline{a_{zz}^{\mathrm{D}} (t)}^{\infty,\omega} & = \frac{2}{25} J
  (\omega) (1  + 2 n (\omega)), \nonumber\\
\overline{a_{z +}^{\mathrm{D}} (t)}^{\infty,\omega} & =  0,\nonumber\\
\overline{a_{++}^{\mathrm{D}}(t)}^{\infty,\omega} & =  \frac{\alpha}{\beta \Lambda} .
\end{align}
The Kossakowski coefficients of the Floquet-Lindblad master equation correspond to the damping rates of the reduced system according to Eq.~\eqref{eq:kossa_deg} in the $z$--$y$--$x$ basis are given by
\begin{align}
\label{eq:dumping_rates_deg}
    b^{\mathrm{D}}_{zz} = \frac{2}{25}J(\omega)(1+n(\omega)), \quad b^{\mathrm{D}}_{yy} = 0, \quad b^{\mathrm{D}}_{xx} =\frac{\alpha}{\beta\Lambda}. 
\end{align}
The approximated expression for the coherent contribution can be obtained in a similar way starting from
Eq.~\eqref{eq:hp_deg}, so that we obtain
\begin{equation}
  \overline{h_+^{\mathrm{D}} (t)}^{\infty,\omega} =  0.
\end{equation}

Finally, it is instructive to observe that, according to 
the relations Eq.~\eqref{eq:hzhp}, 
obtaining the Floquet-Lindblad master equation in the non-degenerate case requires performing the secular approximation, which analogously to the degenerate case, amounts to averaging over the oscillations with frequency $2\varepsilon$. This procedure provides an accurate description of the dynamics only when the timescale $\varepsilon^{-1}$ is much shorter than the characteristic timescale of the reduced system dynamics. Consequently, this approximations break down when the quasienergies become nearly degenerate. The failure of the Floquet-Lindblad description in the vicinity of quasienergy degeneracies has been noted in Ref.~\cite{one_over_f_noise,Mozgunov2020}, motivating the development of partial secular approximations.

\subsection{Non-Markovianity in presence and absence of degeneracy}
In order to study the non-Markovianity of the considered open driven system we proceed as follows. We numerically integrate Eq.~\eqref{eq:dist_dyn} for the distance vector and then compute the non-Markovianity according to Eq.~\eqref{eq:NM_bloch} by maximizing over a thousand randomly sampled vectors, each corresponding to an orthogonal pair of states. 
This evaluation is performed for values of the  driving strength $\Omega$ up to $10\omega_0$.
For simplicity, the simulations are performed at exact resonance, that is $\omega= \omega_0$. The parameters of the 
spectral density are chosen so as to ensure that the system-bath coupling is weak enough so that the Floquet-TCL approximation holds, with a cut-off frequency comparable to the system frequency, so as to allow for non-Markovian effects.
\begin{figure}[t]
    \centering
    \includegraphics[width=\columnwidth]{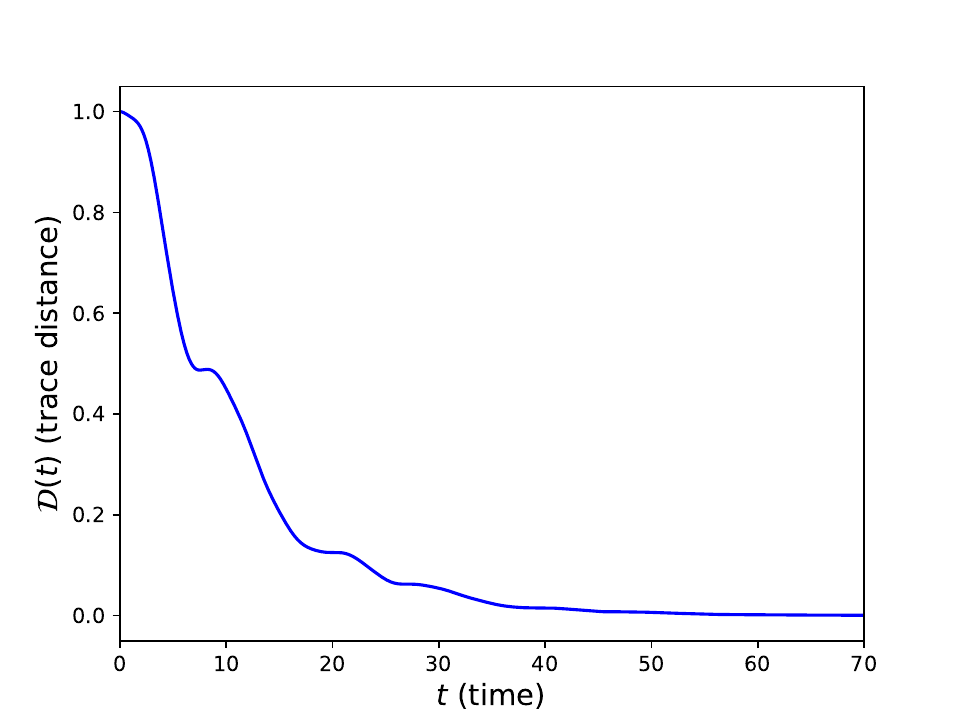}    
    \vspace{0.2cm}    
    \includegraphics[width=\columnwidth]{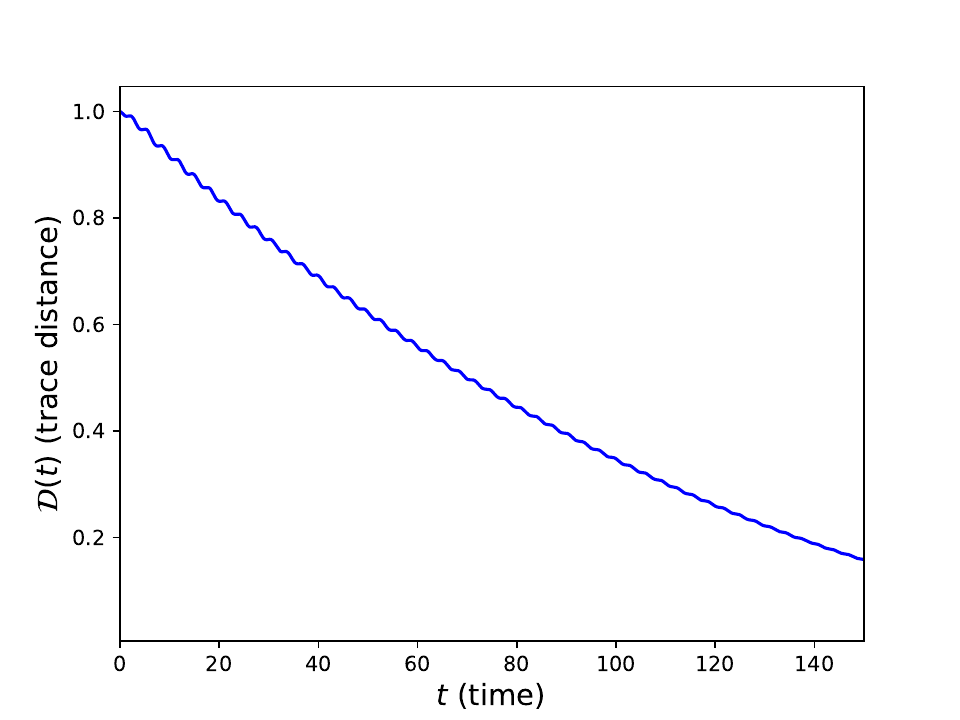}    
    \caption{Top: Trace distance for a pair of states maximizing the non-Markovianity in the non-degenerate case, corresponding to $\boldsymbol{d}_0 = (-0.25,   0.88,  0.39)$ for $\varepsilon = 0.167$. Bottom: Trace distance for a pair of state maximizing non-Markovianity in the degenerate case, that is  $\varepsilon = 0$. Now the optimal pair corresponds to the vector $\boldsymbol{d}_0 =  (0.99,  0.02, -0.11)$. In both cases with consider a  Lorentz-Drude spectral density as in Eq.~\eqref{LD_Spectral} with $\alpha=0.1\omega_0$, $\Lambda=1.2\omega_0$ and $\beta=1.0\omega_0$.}
    \label{fig:trace_distances}
\end{figure}
In Fig.~\ref{fig:trace_distances}(top), the time evolution of the trace distance for the pair of orthogonal states maximizing the non-Markovianity is shown for the non-degenerate case. 
The trace distance rapidly decays to zero, allowing only a few revivals to occur during the dynamics.
Figure~\ref{fig:trace_distances}(bottom) shows the corresponding result for the degenerate case, where the optimal pair of initial states as expected is almost perfectly aligned with the $x$ direction. Indeed, for the parameters specified above, it is straightforward to verify that $J(\omega)(2N(\omega)+1)\approx\alpha/(\beta\Lambda)$, from which Eq.~\eqref{eq:dumping_rates_deg} immediately implies that $b^{\mathrm{D}}_{zz}\ll b^{\mathrm{D}}_{xx}$. This, together with the form of the dynamical matrix in the degenerate case [Eq.~\eqref{eq:dynamical_deg}], implies that the relaxation time associated with the $x$ direction, which is proportional to $1/b^{\mathrm{D}}_{zz}$, is significantly enhanced, thereby allowing for persistent and repeated trace distance revivals.
We note that, in the parameter regime considered here, the $x$ direction in the Floquet basis is less affected by decoherence. At the quasienergy degeneracies, this direction decouples from the $y$ direction and remains only weakly coupled to the $z$ direction, thereby providing an almost protected subspace, we call this mechanism \emph{quasienergy-induced dissipative decoupling}.

\begin{figure}[t]
    \centering
    \includegraphics[width=\columnwidth]{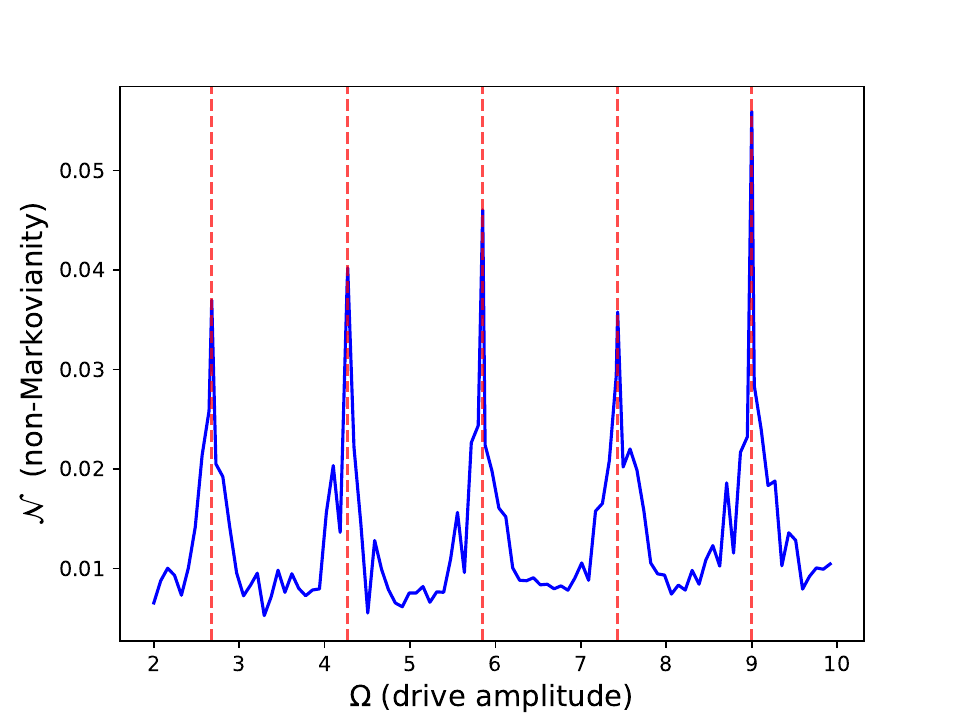}    
    \vspace{0.2cm}    
    \includegraphics[width=\columnwidth]{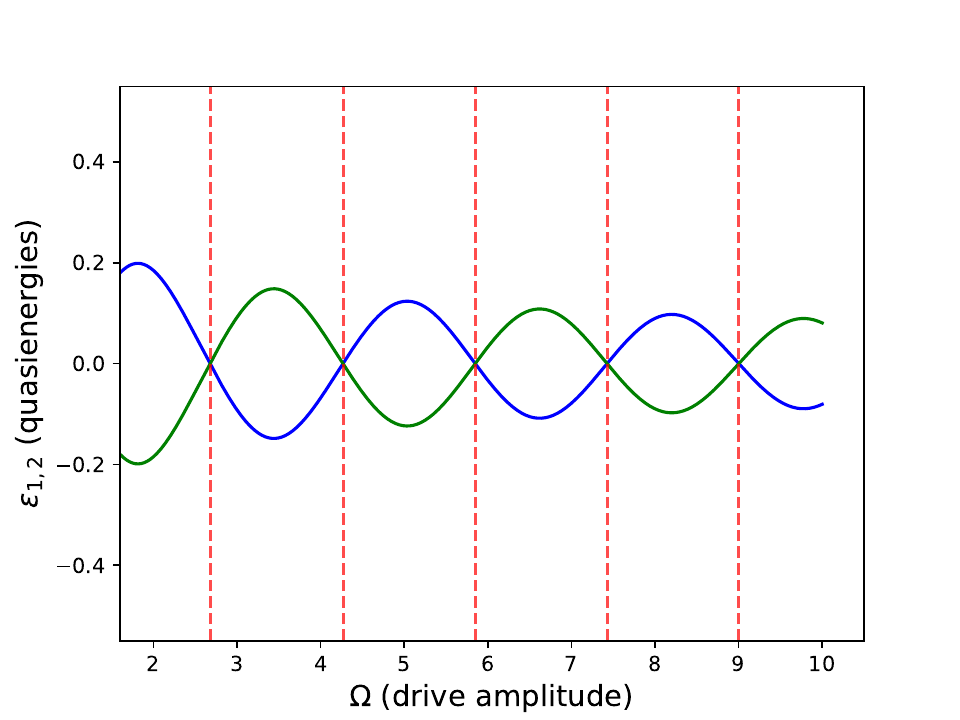}    
    \caption{Non-Markovianity (top) and quasienergies (bottom) as a function of the driving strength $\Omega$. Vertical dashed red lines indicate quasienergy degeneracies, occurring at $\Omega = \{2.68, 4.27, 5.88, 7.43, 9.00\}\omega_0$. The relevant parameters of the spectral density are the same as in Fig.~\ref{fig:trace_distances}. }
    \label{fig:peaks_NM}
\end{figure}

Furthermore, it is important to note that, in order to obtain a Markovian master equation, it is necessary to average over the period of the driving. This leads to neglecting the trace distance revivals which are, as can be seen in Fig.~\ref{fig:trace_distances}(bottom), rapidly oscillating with respect to the dynamics of the reduced system. Finally, we evaluate non-Markovianity as a function of the driving amplitude $\Omega$ within the regime of validity of the approximation for the coefficients given in Eq.~\eqref{eq:coeff_linearly_driven}. The results are shown in Fig.~\ref{fig:peaks_NM}(top), while Fig.~\ref{fig:peaks_NM}(bottom) shows the corresponding quasienergies. Sharp and well-defined peaks are observed at the quasienergy degeneracies (indicated by the vertical dashed lines), as a direct consequence of the mechanism discussed above. This shows that the Floquet-TCL master equation, despite being a second-order approximation, accurately faithfully captures the non-Markovian effects of the dynamics. These peaks have been previously observed in Ref.~\cite{Follia2026a} relying on numerically exact HEOM method, while their connection with longer relaxation times along a given direction in the presence of quasienergy degeneracies was figured out, relying on a Markovian model described by a Floquet-Lindblad master equation. However, that master equation predicts no memory effects at all. We have now provided a microscopic analytical treatment capable of actually describing and clarifying the physical interpretation of the peaks. The agreement with the non-Markovianity peaks obtained from exact numerical simulations based on the HEOM method~\cite{Follia2026a} is striking. This result, on the one hand, confirms the validity of the present approach in that it is able to reproduce the results obtained by numerically exact simulations involving all degrees of freedom; on the other hand, it grounds the previously proposed interpretation on a truly non-Markovian dynamical model.
\section{Conclusions}
\label{sec:ceo}
We have derived the Floquet-TCL master equation for periodically driven open two-level systems by combining the second-order time-convolutionless formalism with Floquet theory to decompose the system coupling operator in the Floquet basis~\cite{Blumel1991a,Breuer2002}. 
This master equation proves particularly convenient for capturing the effects of quasienergy degeneracies on the dynamical properties of the system.

Indeed, we have shown that, under suitable assumptions or approximations, quasienergy degeneracies induce a significant structural modification of the dissipative contribution. In particular, the dissipative dynamics becomes restricted to a two-dimensional subspace of the Bloch sphere, while the coherent contribution is confined to the direction orthogonal to this dissipative plane, a mechanism we called quasienergy-induced dissipative decoupling. This structural change can have profound implications for the dynamics of the system.

After deriving the Floquet-TCL master equation for different relevant realizations of the damped two-level system, we investigated the effects of quasienergy degeneracies by studying the trace-distance-based measure of non-Markovianity for the linearly driven spin-boson model. We found sharp and well-defined peaks in the non-Markovianity at the quasienergy degeneracies. Such peaks were previously observed in Ref.~\cite{Follia2026a} using the numerically exact hierarchical equations of motion, and the agreement with those results is striking. On the one hand, this demonstrates that the present approach accurately reproduces the results of numerically exact simulations. On the other hand, it provides a physical explanation of this phenomenon in terms of the structural modification of the master equation at quasienergy degeneracies.

The emergence of peaks in the non-Markovianity, as discussed in Sec.~\ref{sec:nm}, is directly connected to the appearance of an almost decoherence-protected subspace through the quasienergy-induced dissipative decoupling mechanism.  This feature was already identified in Ref.~\cite{Follia2026a} within a Markovian description and is here confirmed by the Floquet-TCL master equation.

These findings can have important implications for coherence-protection protocols and provide complementary insights to previous works.
The master equation derived in this work provides a versatile framework for describing memory effects in open quantum systems. It further paves the way for the investigation of non-Markovian dynamics in a broad class of scenarios, including different system-environment coupling mechanisms, arbitrary environmental spectral densities, and general forms of periodic driving.

\appendix 
\section{Correlation Function}
\label{Appendix:Correlation Function}
In this appendix, we derive an explicit expression for the bath correlation function starting from Eq.~\eqref{eq:Fluct_diss_theo}
where the spectral density is of the Lorentz-Drude form  [Eq.~\eqref{LD_Spectral}].
We first use the identity~\cite{Grad1963a}
\begin{equation}
    \left( \frac{a}{\pi} \right)^2 \sum_{k = 1}^{\infty} 
    \frac{1}{( {a}/{\pi})^2 + k^2}   =  \frac{1}{2} (a \coth (a) - 1)
\end{equation}
to express the hyperbolic cotangent contribution in the integrand as
\begin{equation}
\label{eq:exp_coth}
    \coth \left( \frac{\beta \varpi}{2} \right)=\frac{2}{\beta \varpi}+\frac{4}{\beta \varpi} \sum_{k = 1} \frac{\varpi^2}{\varpi^2 + \nu_k^2},
\end{equation}
where we have introduced the Matsubara frequencies
\begin{align}
  \nu_k & =  \frac{2 \pi}{\beta} k,
\end{align}
with $k \in \mathbb{N}$.
We can thus evaluate the integrals for each fixed $k$ by means of the residue theorem and resum some of the obtained contributions using the companion identity~\cite{Grad1963a}
\begin{equation}
    \left( \frac{a}{\pi} \right)^2 \sum_{k = 1}^{\infty} \frac{1}{({a}/{\pi})^2 - k^2}   =  \frac{1}{2} (a \cot (a) - 1),
\end{equation}
implying in particular
\begin{equation}
      \label{eq:exp_cot}
   \sum_{k = 1} \frac{\Lambda^2}{\Lambda^2 - \nu_k^2} = \frac{\beta \Lambda}{4} \cot \left( \frac{\beta \Lambda}{2} \right) - \frac{1}{2}.
\end{equation}
We thus finally obtain the compact expressions
\begin{align}
\label{eq:explicit_corr_function}
  \mathrm{Im}\{C (t)\}& = -\alpha \frac{\Lambda}{4}  \mathrm{e}^{- \Lambda t}, \nonumber \\
  \mathrm{Re} \{C (t)\} & =  \alpha \frac{\Lambda}{4} \cot \left( \frac{\beta
  \Lambda}{2} \right) \mathrm{e}^{- \Lambda t} - \frac{\alpha}{\beta \Lambda} \sum_{k
  = 1} \nu_k \frac{\Lambda^2}{\Lambda^2 - \nu_k^2} \mathrm{e}^{- \nu_k t} .
\end{align}

\section{Explicit expression of coefficients}
\label{Appendix:explicit_coefficients}
We now explicitly compute the Kossakowski coefficients and the coherent contribution appearing in the Floquet-TCL master equation Eq.~\eqref{eq:Floquet_TCL_zpm} for the considered linearly driven spin-boson model, relying on their definition relations Eqs.~\eqref{eq:Kossa_zpm} and~\eqref{eq:LS_Hamiltonian}. This is achieved by employing the approximate expressions for  $c_{z}(t)$ and $c_{\pm}(t)$ given in Eq.~\eqref{eq:coeff_linearly_driven}, together with the explicit form of the bath correlation function derived in Appendix~\ref{Appendix:Correlation Function}, and substituting them in Eqs.~\eqref{eq:gamma} and~\eqref{eq:pi}. 
As a result all coefficients can be expressed in terms of the integrals
\begin{align}
  I_1 &=  \int_0^t d s \cos (\omega s) \mathrm{Re} \{C (t - s)\},\nonumber\\
  I_2 &=  \int_0^t d s \cos (\omega s) \mathrm{Im}\{ C (t - s)\}, \nonumber \\
  I_3 &=  \int_0^t d s \,\mathrm{e}^{- i 2 \varepsilon s} \mathrm{Re} \{C (t - s)\}, \nonumber \\
  I_4 &=  \int_0^t d s \,\mathrm{e}^{- i 2 \varepsilon s} \mathrm{Im} \{C (t - s)\},
\end{align}
which can be evaluated explicitly using Eq.~\eqref{eq:explicit_corr_function}, yielding
\begin{align}
\label{eq:rel_int}
  I_1  = &  \frac{1}{4} \left\{
  - \cot \left( \frac{\beta \Lambda}{2} \right) \frac{J (\omega)}{\omega}
  \Lambda \mathrm{e}^{- \Lambda t} \right. \nonumber\\
  &  + J (\omega) \left[ \coth \left( \frac{\beta \omega}{2} \right) \cos
  (\omega t) + \cot \left( \frac{\beta \Lambda}{2} \right) \sin (\omega t)
  \right] \nonumber\\
  &  \left. + \frac{4 \alpha}{\beta \Lambda} \sum_{k = 1}
  \frac{\Lambda^2}{\Lambda^2 - \nu_k^2}  \frac{\nu_k}{\omega^2 +
  \nu_k^2} [\nu_k \mathrm{e}^{- \nu_k t} - \omega \sin (\omega t)] \right\},
  \nonumber\\
  I_2 = &  \frac{1}{4} \left\{
  \frac{J (\omega)}{\omega} \Lambda \mathrm{e}^{- \Lambda t} - \frac{J
  (\omega)}{\omega} \cos (\omega t) - J (\omega) \sin (\omega t) \right\}, \nonumber\\ 
  I_3 = &  \frac{
  \alpha}{4} \left\{ \cot \left( \frac{\beta \Lambda}{2} \right)
  \frac{\Lambda}{\Lambda - i 2 \varepsilon} (\mathrm{e}^{- i 2 \varepsilon t} -
  \mathrm{e}^{- \Lambda t}) \right. \nonumber\\
  &  \left. - \frac{4}{\beta \Lambda} \sum_{k = 1}
  \frac{\Lambda^2}{\Lambda^2 - \nu_k^2} \frac{\nu_k}{\nu_k - i 2
  \varepsilon} (\mathrm{e}^{- i 2 \varepsilon t} - \mathrm{e}^{- \nu_k t}) \right\},
  \nonumber\\
  I_4 = &  \frac{\alpha}{4} \left\{ \frac{\Lambda}{\Lambda - i 2 \varepsilon} (\mathrm{e}^{- i 2
  \varepsilon t} - \mathrm{e}^{- \Lambda t}) \right\}.
\end{align}
\begin{widetext}
Using the auxiliary integrals defined above, the explicit expression of the Kossakowski coefficients~\eqref{eq:Kossa_zpm} becomes
\begin{align}
    a_{zz} (t) = & \frac{4}{25} \cos (\omega t) \left\{ J (\omega) \left[ \coth \left( \frac{\beta \omega}{2} \right) \cos (\omega t) + \cot \left( \frac{\beta \Lambda}{2} \right) \left( \sin (\omega t) - \frac{\Lambda}{\omega} \mathrm{e}^{- \Lambda t} \right) \right] \right. \nonumber \\
    & \left. + \frac{4 \alpha}{\beta \Lambda} \sum_{k = 1} \frac{\Lambda^2}{\Lambda^2 - \nu_k^2} \frac{\nu_k}{\omega^2 + \nu_k^2} \left[ \nu_k \mathrm{e}^{- \nu_k t} - \omega \sin (\omega t) \right] \right\},
    \label{eq:coeff_zz}
\\
    a_{z +} (t) = & \frac{1}{5\sqrt{2}} \left\{ \frac{J (2\varepsilon)}{2\varepsilon}({\Lambda + i 2 \varepsilon}) \cos (\omega t) \left( i + \cot \left( \frac{\beta \Lambda}{2} \right) \right) (\mathrm{e}^{- i 2 \varepsilon t} - \mathrm{e}^{- \Lambda t}) \right. \nonumber \\
    & + J (\omega) \mathrm{e}^{- i 2 \varepsilon t} \left[ \frac{\Lambda}{\omega} \left( i - \cot \left( \frac{\beta \Lambda}{2} \right) \right) \mathrm{e}^{- \Lambda t} - \left( i - \cot \left( \frac{\beta \Lambda}{2} \right) \right) \sin (\omega t) - \left(i\frac{\Lambda}{\omega} - \coth \left( \frac{\beta \omega}{2} \right) \right) \cos (\omega t) \right] \nonumber \\
    & \left. + \frac{4 \alpha}{\beta \Lambda} \sum_{k = 1} \frac{\Lambda^2}{\Lambda^2 - \nu_k^2} \nu_k\left[ \frac{ \nu_k \mathrm{e}^{- \nu_k t} - \omega \sin (\omega t) }{\nu_k^2 + \omega^2}\mathrm{e}^{- i 2 \varepsilon t} + \frac{ \mathrm{e}^{- \nu_k t}-\mathrm{e}^{- i 2 \varepsilon t} }{\nu_k - i 2 \varepsilon} \cos (\omega t)\right] \right\},
    \label{eq:coeff_zp}
\end{align}
together with
\begin{align}
\label{eq:coeff_pppm}
a_{+ +} (t) = 
& \frac{1}{2} \left\{  J (2 \varepsilon) (\coth (\beta
  \varepsilon) - 1) 
  +  J (2 \varepsilon) \left[ \left(\cot \left(
  \frac{\beta \Lambda}{2} \right)+\frac{\Lambda}{2 \varepsilon }\right)\sin (2 \varepsilon t) +
   \left(1-\frac{\Lambda}{2 \varepsilon }\cot \left(
  \frac{\beta \Lambda}{2} \right)\right)\cos (2 \varepsilon t) \right]  \mathrm{e}^{- \Lambda t} 
  \right. \nonumber\\
  &   \left. + \frac{4\alpha}{\beta \Lambda} \sum_{k = 1}
  \frac{\Lambda^2}{\Lambda^2 - \nu_k^2}  \frac{\nu_k}{\nu_k^2 + 4
  \varepsilon^2} [\nu_k \cos (2 \varepsilon t) - 2 \varepsilon \sin (2
  \varepsilon t)] \mathrm{e}^{- \nu_k t} \right\} ,
    \\ 
    a_{+ -} (t) = &\mathrm{e}^{i 2 \varepsilon t} \left\{ \frac{1}{2} \cot \left( \frac{\beta \Lambda}{2} \right) \frac{J (2 \varepsilon)}{2 \varepsilon} (\Lambda - i 2 \varepsilon) (\mathrm{e}^{i 2 \varepsilon t} - \mathrm{e}^{- \Lambda t}) - \frac{2 \alpha}{\beta \Lambda} \sum_{k = 1} \frac{\Lambda^2}{\Lambda^2 - \nu_k^2} \frac{\nu_k}{\nu_k + i 2 \varepsilon} (\mathrm{e}^{i 2 \varepsilon t} - \mathrm{e}^{- \nu_k t}) \right\}.
\end{align}
These coefficients fix also the remaining ones according to the relations $ a_{z -} (t) =   a_{z +} (t)_{| \varepsilon \rightarrow - \varepsilon}
$ and $ a_{- -} (t) =   a_{+ +} (t)_{| \varepsilon \rightarrow - \varepsilon}$.
Finally, the Lamb-shift Hamiltonian~\eqref{eq:LS_Hamiltonian} is expressed in terms of the coefficients
\begin{align}
\label{eq:hzhp}
    h_z (t) = & \frac{1}{2 \sqrt{2}} \left\{  J (2 \varepsilon) \left[ 1-\left(\cos (2 \varepsilon t) + \frac{\Lambda}{2 \varepsilon} \sin (2 \varepsilon t) \right)\mathrm{e}^{- \Lambda t}\right] \cot \left( \frac{\beta \Lambda}{2} \right)  \right. \nonumber \\
    & \left. - \frac{4 \alpha}{\beta \Lambda} \sum_{k = 1} \frac{\Lambda^2}{\Lambda^2 - \nu_k^2} \frac{2 \varepsilon\nu_k}{4 \varepsilon^2 + \nu_k^2} \left[1 - \left(\cos (2 \varepsilon t) +\frac{\nu_k}{2 \varepsilon}  \sin (2 \varepsilon t)\right) \mathrm{e}^{- \nu_k t} \right]\right\},
\nonumber\\
    h_+ (t) = & \frac{1}{5} \left\{ \cot \left( \frac{\beta \Lambda}{2} \right) \frac{J (2 \varepsilon)}{2 \varepsilon} (\Lambda - i 2 \varepsilon) (\mathrm{e}^{i 2 \varepsilon t} - \mathrm{e}^{- \Lambda t}) \cos (\omega t) \right. \nonumber \\
    & - \cot \left( \frac{\beta \Lambda}{2} \right) J (\omega) \frac{\Lambda}{\omega} \mathrm{e}^{- \Lambda t} \mathrm{e}^{i 2 \varepsilon t} + J (\omega) \left[ \coth \left( \frac{\beta \omega}{2} \right) \cos (\omega t) + \cot \left( \frac{\beta \Lambda}{2} \right) \sin (\omega t) \right] \mathrm{e}^{i 2 \varepsilon t} \nonumber \\
    & \left. - \frac{4 \alpha}{\beta \Lambda} \sum_{k = 1} \frac{\Lambda^2}{\Lambda^2 - \nu_k^2} \frac{\nu_k}{\nu_k + i 2 \varepsilon} (\mathrm{e}^{i 2 \varepsilon t} - \mathrm{e}^{- \nu_k t}) \cos (\omega t) + \frac{4 \alpha}{\beta \Lambda} \sum_{k = 1} \frac{\Lambda^2}{\Lambda^2 - \nu_k^2} \frac{\nu_k}{\omega^2 + \nu_k^2} [\nu_k \mathrm{e}^{- \nu_k t} - \omega \sin (\omega t)] \mathrm{e}^{i 2 \varepsilon t} \right\}.
\end{align}
It is now possible to obtain the Kossakowski coefficients and the Lamb shift Hamiltonian in the degenerate case by simply setting $\varepsilon = 0$ in the expressions derived above, namely
\begin{align}
\label{eq:azp_dege}
    a_{z \pm}^{\mathrm{D}} (t) = & \frac{1}{5\sqrt{2}} \left\{ \alpha\cos (\omega t) \left(i + \cot \left( \frac{\beta \Lambda}{2} \right) \right) (1 - \mathrm{e}^{- \Lambda t}) + 2\alpha \left( \cot \left( \frac{\beta \Lambda}{2} \right) - \frac{2}{\beta \Lambda} \right) \cos (\omega t) \right. \nonumber \\
    & + J (\omega) \left[ \frac{\Lambda}{\omega} \left( i - \cot \left( \frac{\beta \Lambda}{2} \right) \right) \mathrm{e}^{- \Lambda t} - \left( i - \cot \left( \frac{\beta \Lambda}{2} \right) \right) \sin (\omega t) - \left(i\frac{\Lambda}{\omega} - \coth \left( \frac{\beta \omega}{2} \right) \right) \cos (\omega t) \right] \nonumber \\
    & \left. + \frac{4 \alpha}{\beta \Lambda} \sum_{k = 1} \frac{\Lambda^2}{\Lambda^2 - \nu_k^2} \left[ \frac{\nu_k (\nu_k \mathrm{e}^{- \nu_k t} - \omega \sin (\omega t))}{\nu_k^2 + \omega^2} + \mathrm{e}^{- \nu_k t} \cos (\omega t) \right] \right\}
\end{align}
and 
\begin{align}
        \label{eq:appapm_dege}
    a_{\pm \pm}^{\mathrm{D}} (t)   = &  
  \frac{\alpha}{\beta \Lambda} - \frac{\alpha}{2} \cot \left( \frac{\beta \Lambda}{2}
  \right)\mathrm{e}^{- \Lambda t}  + \frac{2\alpha}{\beta \Lambda} \sum_{k = 1}
  \frac{\Lambda^2}{\Lambda^2 - \nu_k^2} \mathrm{e}^{- \nu_k t} ,
\end{align}
while we note that since $a_{zz}(t)$ does not depend on $\varepsilon$, it is not affected by the degeneracy. 
For the 
coherent part, in the degenerate case  we obtain $h_z^{\mathrm{D}} (t) =  0$ together with
\begin{align}
\label{eq:hp_deg}
    h_+^{\mathrm{D}} (t) = & \frac{1}{5} \left\{ \left[ \frac{2\alpha}{\beta \Lambda} - \alpha\cot \left( \frac{\beta \Lambda}{2} \right) \mathrm{e}^{- \Lambda t} \right] \cos (\omega t) - \cot \left( \frac{\beta \Lambda}{2} \right) J (\omega) \frac{\Lambda}{\omega} \mathrm{e}^{- \Lambda t} \right. \nonumber \\
    & \left. + J (\omega) \left[ \coth \left( \frac{\beta \omega}{2} \right) \cos (\omega t) + \cot \left( \frac{\beta \Lambda}{2} \right) \sin (\omega t) \right] + \frac{4 \alpha}{\beta \Lambda} \sum_{k = 1} \frac{\Lambda^2}{\Lambda^2 - \nu_k^2} \left[ \mathrm{e}^{- \nu_k t} \cos (\omega t) + \frac{\nu_k (\nu_k \mathrm{e}^{- \nu_k t} - \omega \sin (\omega t))}{\omega^2 + \nu_k^2} \right] \right\}.
\end{align}
\end{widetext}
\bibliography{biblio,ref}
\end{document}